\documentclass[twocolumn,nofootinbib]{revtex4}
\usepackage{epsfig}
\usepackage{amsmath}
\begin{document}

\title{Cross-fertilization of QCD and statistical physics:\\
high energy scattering, reaction-diffusion, selective evolution, 
spin glasses \\
and their connections\footnote{
Lectures given at the 
Cracow School of Theoretical Physics, XLVI Course,
June 2006, Zakopane, Poland.}
}
\author{St\'ephane Munier
}
%\address{
\affiliation{
Centre de Physique Th\'eorique,
\'Ecole Polytechnique, CNRS, 91128 Palaiseau, France}
%}
%\maketitle

\begin{abstract}
High energy scattering was recently shown to be 
similar to a reaction-diffusion process.
The latter  defines a 
wide universality class
that also contains e.g. some specific population evolution models.
The common point of all these models
is that their respective dynamics
are described by noisy traveling wave equations.
This observation has led to a new understanding of QCD in the regime
of high energies, and
known universal results on reaction-diffusion models could be transposed
to obtain quantitative properties of QCD amplitudes. 
Conversely,
new general results for that kind of statistical 
models have also been derived.
Furthermore,
an intriguing relationship between noisy traveling wave equations
and the theory of spin glasses was found.
\end{abstract}
%\PACS{PACS numbers come here}
\maketitle

\section{Introduction}

The behavior of hadronic cross sections at high energy has been the
subject of
intense theoretical and experimental investigations for several decades.
These studies will become even more relevant with the advent of the LHC,
where protons will collide at a center-of-mass energy of 14~TeV.

In the following, we will focus on the forward elastic
amplitude $A(Y,k)$ for the scattering of two hadronic 
objects at total rapidity $Y$, one of
the objects (called the probe) being 
characterized by a tunable momentum $k$,
that defines the scales in the plane transverse to the
collision axis. 
The other hadron will be called the target.
The scattering occurs at a given impact parameter that, here, will
be fixed.
Experimentally, the corresponding process could be,
for example, deep-inelastic electron-proton (or nucleus)
scattering, in which case $k$ would be of order of
the virtuality of the exchanged photon.
The total rapidity $Y$ can be seen as the rapidity
of the target in the rest frame of the probe: It is related to the squared
center-of-mass energy $s$ through $Y=\ln (s/k^2)$.
Note that one could equally well work in transverse coordinates 
instead of momenta,
which proves useful in many cases: 
In coordinate space, $\tilde A(Y,r)$ is directly related to the
probability that the probe of size $r$ interact with the evolved target
at the considered impact parameter, which in particular results in the
unitarity bound $\tilde A\leq 1$.
The relationship between the two representations of the amplitude
reads
\begin{equation}
\tilde A(Y,r)=r^2\int \frac{d^2 k}{2\pi} e^{i k\cdot r}A(Y,k).
\label{transfo}
\end{equation}
Although total cross sections are not computable from standard
QCD methods since confinement effects necessarily enter,
the evolution with rapidity of the scattering amplitudes 
for a localized probe at
a given impact parameter (understood in Eq.~(\ref{transfo}))
may be obtained starting 
from perturbative
calculations that lead e.g. to the Balitsky hierarchy of
equations \cite{B}.
In the latter framework, the evolution of $\tilde A$ reads
\begin{multline}
\partial_{\bar\alpha Y} \tilde A(Y,r)
=\int \frac{d^2z}{2\pi}\frac{r^2}{z^2(r-z)^2}
(\tilde A(Y,z)+\tilde A(Y,r-z)\\
-\tilde A(Y,r)
-\langle\tilde T(Y,z)\tilde T(Y,r-z)\rangle),
\label{B}
\end{multline}
where $\bar\alpha=\frac{\alpha_s N_c}{\pi}$.
One had to introduce $\tilde T$ which is meant to be
the scattering amplitude off 
{\it one particular configuration} of the
target. It is its average over 
all quantum fluctuations of the latter that corresponds 
to the physically measurable amplitude:
 $\tilde A=\langle \tilde T\rangle$.   

Eq.~(\ref{B}) is not closed:
One needs a further evolution equation for $\langle \tilde
T\tilde T\rangle$.
Eq.~(\ref{B}) is only the first equation of
an infinite hierarchy which involves
increasingly complex mathematical objects. Solving 
that system is a formidable task, and not surprisingly,
no one has found a solution from a direct study 
of Eq.~(\ref{B}). Another problem with the hierarchy is that
it is not clear whether its complete form for arbitrary targets
is known.

A significantly simpler equation may be obtained by arbitrarily
factorizing the correlator 
$\langle\tilde T\tilde T\rangle$ into the product
$\tilde A\tilde A$, 
within a kind of mean field approximation that neglects
fluctuations in the target.
This was first proposed by Balitsky \cite{B}
and subsequently re-derived by Kovchegov \cite{K} 
in the particular physical context
of deep-inelastic scattering off an infinitely large nucleus.
An elegant representation of the resulting equation
is obtained in momentum space 
by using Eq.~(\ref{transfo}), namely
\begin{equation}
\partial_{\bar\alpha Y} A=\chi(-\partial_{\ln k^2})A-A^2,
\label{BK}
\end{equation}
where $\chi(-\partial_{\ln k^2})$ 
is the expression of the integral kernel that appears
in Eq.~(\ref{B}) in momentum space.
The main properties of the solutions to Eq.~(\ref{BK}) are now 
known~\cite{LT,MT2002,MP2003,MP2004}, but 
the effects of the fluctuations neglected in going to the
Balitsky-Kovchegov (BK) equation~(\ref{BK}) 
and even their very nature had not been appreciated until 
quite recently.

One may guess 
that one difficulty with Eqs.~(\ref{B}) and~(\ref{BK}) 
is that they both contain nonlinearities.
The latter are meant 
to encode parton saturation effects~\cite{GLRMQ}.
One could think of neglecting them
(by dropping the $-A^2$ term in Eq.~(\ref{BK}) for
example), and indeed, that used to be
the usual approximation until a few years ago.
The resulting equation is named after 
Balitsky, Fadin, Kuraev and Lipatov (BFKL) 
\cite{BFKL}
($\chi(-\partial_{\ln k^2})$ that
appears in Eq.~(\ref{BK}) is called the BFKL kernel).
But in 1999, Golec-Biernat and W\"usthoff showed \cite{GBW} that
the nonlinearities in Eqs.~(\ref{B}),(\ref{BK})
may already play an important role in the
most recent data for deep-inelastic scattering
(see also Ref.~\cite{MSM}). Their work stirred a great theoretical 
interest for
the full nonlinear equations.
For a long time,
it was the BK equation that was the focus of most of the theoretical
and phenomenological works in high energy QCD, despite the relative
arbitrariness of the approximations made to get it.
Only in 2004 did
Mueller and Shoshi address the problem
of trying to quantify
effects beyond those described by the BK equation,
using a very original approach \cite{MS2004}.
Subsequently, a deep interpretation of what they had done
in relation with problems of statistical mechanics
was found \cite{M2005,MP2004,EGBM2005}, 
which paved the way for a new
and fruitful understanding of high energy QCD.

The goal of these two lectures is to show 
that high energy scattering looks very much like some processes
(called reaction-diffusion)
that appear in other physical, biological, or chemical contexts.
We will provide a very simple and transparent picture of high energy
QCD, but accurate enough to enable one to get presumably
exact analytical results for QCD.
The first lecture (Sec.~\ref{lecture1}) aims at establishing this
correspondence.
``Cross-fertilization'' is the title of these lectures because while
investigating this relationship, we were able to also derive some
general results that apply to a wider context. 
The second lecture (Sec.~\ref{lecture2}) reviews them.
The most important
of these results may directly be transposed to QCD.
Throughout, the focus will be put on the underlying 
physical ideas rather than
on the technicalities (references to original
work where details are worked out will be provided whenever 
necessary).
We would like the reader to appreciate that
this business was made possible by coming back to the
basic concepts of high energy scattering in the parton model
in the light of apparently completely unrelated physics,
rather than trying to address directly the very technical 
equations that
have been established for QCD.

%%%%%%%%%%%%%%%%%%%%%%%%%%%%%%%%%%%%%%%%%%%%%%%%%%%%%%%%%%%

\section{\label{lecture1}High energy QCD as a statistical process}

\subsection{What is universality?}

Our first task is to understand that it may sometimes
be fruitful to replace
a complicated problem (like QCD evolution, see Eq.~(\ref{B}))
by a much simpler one (reaction-diffusion in our case, as we will
see in these lectures), 
and yet be able to get {\it quantitative}
results from the solution to the latter. This is possible
when both models, the simple and the complicated ones, belong
to the same {\it universality class}.

The concept of
universality was introduced in the 60's by a bunch of
renowned physicists, among whom Widom, Kadanoff, Fisher and Wilson 
in the theory of critical phenomena.
To illustrate it, let us consider a system of spins, 
that is a set of binary variables $S_i=\pm 1$, on a two-dimensional lattice
and interacting through the Hamiltonian
\begin{equation}
H=-\sum_{i,j}J_{ij}S_i S_j.
\label{Hising}
\end{equation}
If $J_{ij}=J>0$ when the sites $i$ and $j$ are nearest neighbors and 0 otherwise,
this is the Ising model with ferromagnetic interactions.
The partition function is given by
\begin{equation}
Z=\sum_{\{S_i\}} e^{-H[\{S_i\}]/k_B T}
\label{Zising}
\end{equation}
at a given temperature $T$.

At high temperature, all spin configurations have equal probability.
At low temperature, only a few configurations have a significant probability,
the ones that minimize the energy. The minimum energy configuration is
the one in which all spins are aligned.
Near the critical point, the average magnetization reads
\begin{equation}
\langle S_i\rangle\sim (T_c-T)^{\beta},
\end{equation}
and an exact calculation gives $\beta=1/8$ 
for two-dimensional systems, as may be checked in standard statistical
mechanics textbooks.

The Ising model may represent only a very idealized magnet.
One may wonder why it is interesting to 
study extensively such an over-simplified model, which obviously
incorporates only a few very basic properties of magnetic materials and
neglects many others.
Well, it turns out that the exponent $\beta$ 
(as well as a few other exponents) is
completely independent of the microscopic details of the system, which means
that it is the {\it exact} result also for realistic magnets.
Of course, other quantities strongly depend on those details:
This is the case for the critical temperature $T_c$ for example.

The universality class is defined only by very general properties
of the system, like the space dimension, and the symmetries.
All systems that share these few properties are expected to
bear the same critical exponents. So if one is able to solve
the simplest model, one is likely to know important features 
of all representatives of its universality class.

High energy QCD has little to do with the Ising model and with criticality.
We are now going to introduce a class of systems which, as we
will argue later on, are likely to share universal properties with
high energy QCD.

%%%%%%%%%%%%%%%%%%%%%%%%%%%%%%%%%%%%%%%%%%%%%%%%%%%%%%%%%%%%%%%%%%%%%

\subsection{Warm-up: Properties of a stochastic zero-dimensional model}

We address the time evolution of a system of $n(t)$ particles that obey the
following rules: In each time interval $dt$, each single
particle has a probability $dt$
to split in 2 particles, and a probability $(n(t)-1)dt/N$ to
disappear.
This yields:
\begin{equation}
n(t+dt)=\left\{
\begin{aligned}
n(t)\!+\!1 & \text{  proba  } dt\\
n(t)\!-\!1 & \text{  proba  } \frac{(n(t)\!-\!1)dt}{N}\\
n(t)   & \text{  proba  } 1\!-\!dt -\frac{(n(t)\!-\!1)dt}{N}.
\end{aligned}
\right.
\label{rule0}
\end{equation}
One may represent the evolution of the system in the form of
a stochastic evolution equation:\footnote{
Eq.~(\ref{stocha0}) (as well as all other stochastic equations
that appear in these lectures) has to be interpreted in the It\^o way.
}
\begin{equation}
\frac{dn}{dt}=n-\frac{n(n-1)}{N}+\sqrt{n+\frac{n(n-1)}{N}}\,\nu,
\label{stocha0}
\end{equation}
where $\nu$ is a noise (i.e. a random function of time) of zero average
and of variance satisfying
$\langle \nu(t)\nu(t^\prime)\rangle=\delta(t-t^\prime)$. 
This means that
$\nu$ typically varies by 1 unit when $t$ is changed by 1.
It is a simple exercise to derive Eq.~(\ref{stocha0}): It is enough to compute
the mean $\langle n(t+dt) \rangle$ and variance $\langle n^2(t+dt) \rangle$ 
given $n(t)$ from the stochastic
rules~(\ref{rule0}). 

We see on Eq.~(\ref{stocha0})\footnote{
Note that the form~(\ref{stocha0}) is not particularly useful
in practice, since $\nu$ does not have any special properties.
Our purpose in writing~(\ref{stocha0})
is to get a feeling about the different possible representations
of a stochastic process.
We may also comment that there is a way to 
write the evolution as an It\^o equation 
in which $\nu$ would have
a Gaussian distribution, and this is more interesting
from a technical point of view.
We refer the reader to~\cite{M2006} for the details.
}
that in a typical realization, the number of particles
will start to grow exponentially (linear term) until 
$n(t)\sim N$, at which point
a steady state is reached (nonlinear term), up to 
random fluctuations (noise term).

\begin{figure}
\begin{center}
\epsfig{file=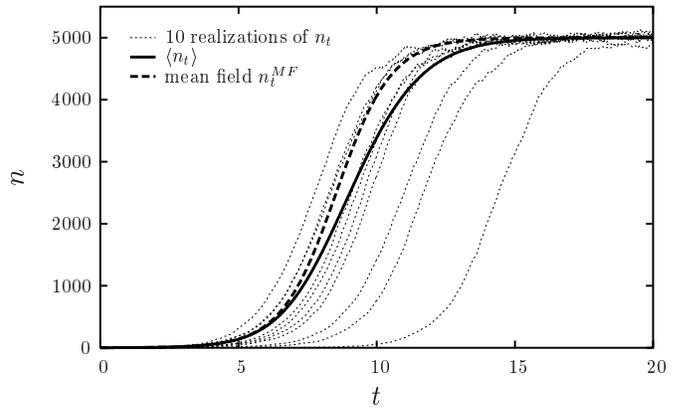,width=9cm}
\end{center}
\caption{\label{fig:zeroD}
(From Ref.~\cite{M2006}).
Ten realizations of the evolution~(\ref{rule0}) of the zero-dimensional 
toy model (dotted lines), together
with the full solution for $\langle n(t)\rangle$ (full line) and
the solution of the mean-field equation~(\ref{mf0}) (dashed line). 
$N$ is set to 5000 in this example.}
\end{figure}

Solving the model consists in exhibiting the statistics of $n(t)$.
In a first approach, one may simulate numerically different 
realizations of the model, and then
perform the average over the realizations
to get $\langle n(t)\rangle$ and maybe higher-order moments of $n(t)$. 
This is shown in Fig.~\ref{fig:zeroD}.
One could also try to obtain an evolution equation for 
$\langle n(t)\rangle$ from
Eq.~(\ref{stocha0}) and solve it analytically. But the latter
involves the correlator 
$\langle n^2\rangle$:
\begin{equation}
\frac{d\langle n\rangle}{dt}=\langle n\rangle 
-\frac{\langle n(n-1)\rangle}{N},
\label{hzero}
\end{equation}
and thus one eventually needs to solve an infinite hierarchy of
coupled equations.
This set of equations is qualitatively
similar to the Balitsky hierarchy~(\ref{B}). 
Only through a mean field approximation, 
consisting in factorizing all correlators and
valid in the large-$N$ limit does Eq.~(\ref{hzero}) 
boil down to a closed equation
\begin{equation}
\frac{d\langle n\rangle}{dt}=\langle n\rangle 
-\frac{\langle n\rangle^2}{N}.
\label{mf0}
\end{equation}
This is the equivalent of the BK equation~(\ref{BK}).
Unfortunately, the solution to this much simpler equation is quite far
from the complete solution to Eq.~(\ref{stocha0}) or~(\ref{hzero})
(see Fig.~\ref{fig:zeroD}),
and it is manifest that more sophisticated approaches to the 
full stochastic model are needed.

To address the complete stochastic model~(\ref{rule0}), 
we can think of two possible ways. One may compute iteratively
the successive orders in $1/N$ by field-theoretical methods
and perform a resummation: This method
was pushed quite far
in Ref.~\cite{SX2005}.
Another possible path 
is to notice that the noise is actually important only
as long as the number of particles $n(t)$ is small: 
For large $n(t)$, the evolution
is essentially deterministic. For small $n(t)$ instead, 
it is the nonlinearity in Eq.~(\ref{stocha0}) that
is negligible, which makes the evolution tractable,
even analytically in this
simple zero-dimensional case.
That was investigated in Ref.~\cite{M2006}.

%%%%%%

\subsection{Universality class of traveling wave  equations}

We now add a spatial dimension to the model.
We consider $n(t,x)$ particles at time $t$ and on each site $x$ of a 
one-dimensional lattice. In addition to the rules~(\ref{rule0}) that apply on
each site, a given particle at $x$ has some probability $p$ to jump to the
nearby sites $x+\Delta x$ or $x-\Delta x$. The evolution of
one realization may then be written as\footnote{
To keep the equation simple and transparent, we replaced the noise term
by its dominant part for $n\ll N$,
up to a constant. In other words, $\nu$ has a variance that
is only approximately normalized to unity. This does not
change the discussion qualitatively.
}
\begin{multline}
\frac{dn(t,x)}{dt}=
p[n(t,x+\Delta x)+n(t,x-\Delta x)-2n(t,x)]\\
+n(t,x)-\frac{n(t,x)(n(t,x)-1)}{N}+\sqrt{n(t,x)}\,\nu,
\label{stocha1}
\end{multline}
where $\nu$ is again a noise of zero mean that varies by typically 1 when $x$ or $t$
are changed by one unit.
This equation is very close to Eq.~(\ref{stocha0}) except for a new
diffusion term
(inside the square brackets) that correlates nearby spatial sites.

Both in Eq.~(\ref{stocha0}) and in Eq.~(\ref{stocha1})
the noise term is of order $\sqrt{n}$. This is 
typical of a statistical noise associated to discrete systems of particles: 
Indeed, adding stochastically
$n$ particles {\it on the average} to a system
means adding a number of particles typically in the range 
$n\pm \sqrt{n}$ {\it in a given realization}.
Hence the noise term is directly related to discreteness.

\begin{figure}
\begin{center}
\epsfig{file=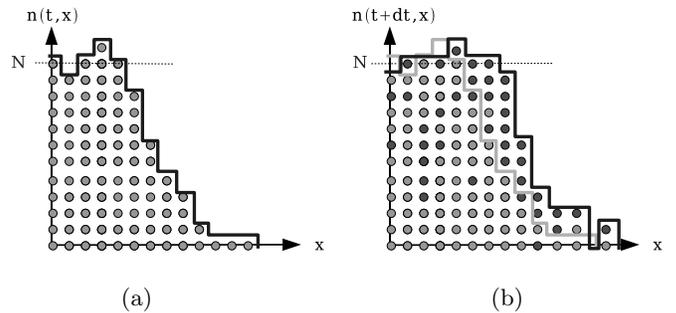,width=9.cm} \\
(a)\hskip 4.5cm  (b)
\end{center}
\caption{\label{fig:rd1}Sketch of one step of the
evolution of a realization 
in the one-dimensional model (from (a) to (b); $N=12$ here).
The dark disks represent the particles that have been added or that have
diffused. For the smallest values of $x$, some particles have also disappeared,
so that the maximum particle number does not exceed $N$ on the average,
up to fluctuations of order $\sqrt{N}$ authorized by the stochastic
term in Eq.~(\ref{stocha1}).
}
\end{figure}

Let us follow a particular realization of the evolution of the system.
Starting from a given initial condition (for example a bunch of 
particles localized
around some given lattice site), 
the number of particles grows exponentially
on each occupied
site until it reaches $N$, at which point the growth stops 
because the positive growth term $n$ is exactly compensated by the negative
nonlinear term $-n^2/N$.
At this point, only fluctuations of order $\sqrt{N}$ are left
in the evolution.
At the same time, diffusion takes place, allowing the particles 
to ``escape'' toward larger
(or lower) values of $x$ (there is a symmetric front traveling to the left if the
initial condition is local). It is then clear that the solution will look like a
noisy wave front, that connects a region (say to the right) where there is no
particle to a region (to the left) where the equilibrium number of particles per site
$N$ is reached. This wave front ``travels'' toward larger values of
$x$, and is therefore
called a {\it traveling wave}. One step of the evolution is 
illustrated in Fig.~\ref{fig:rd1}.

The model that we have outlined is a typical reaction-diffusion
process.
The name ``reaction-diffusion'' stems from chemistry: Such a process
may be a (very crude) model for chemical reactions.
Renormalizing the number of particles through the introduction of $u=n/N$,
one may write the general structure of the evolution of such a system 
in the following form:
\begin{multline}
\partial_t u=\underset{\text{branching diffusion}}{\underbrace{\chi(-\partial_x)u}}
-\underset{\text{compensates the growth of $u$ near 1}}
{\underbrace{\text{nonlinear function of $u$}}}\\
+\underset{\text{encodes discreteness}}
{\underbrace{\text{noise of order} \sqrt{\frac{u}{N}}}}.
\label{evolgen}
\end{multline}
$\chi(-\partial_x)$ is a kernel that encodes branching diffusion. 
It could be a differential operator
such as $\partial_x^2+1$ (which would correspond to carefully performing the 
limit $\Delta x\rightarrow 0$ in the
above example) or, more generally, an integral operator (which
is the case in QCD).

The universality class defined by Eq.~(\ref{evolgen})
is usually named
after Fisher and Kolmogorov-Petrovsky-Piscounov (F-KPP)
\cite{FKPP1937}, 
who for the first time
formulated mathematically such processes.
For a given stochastic equation, there is no general theorem to decide whether
it belongs to the same class as Eq.~(\ref{evolgen})
or not. It is rather a matter of guess from 
the physical properties of the underlying
model. One knowns however that
the exact form of the nonlinearity in Eq.~(\ref{evolgen})
is not important. It could
be any reasonable power of $u$, for example. Also the precise form of the noise
is not an issue and does not affect the properties of the solutions at large $N$: 
One could have a slightly modified form, for example $\sqrt{u(1-u)/N}$. The only
important feature is that it should scale like $\sqrt{u/N}$ 
for small values of $u$.

We are now going to give some basic properties of the solutions to
F-KPP-like evolution equations~\cite{P2004}.
More details 
and a review of the latest developments
will be provided in Sec.~\ref{lecture2}.

\begin{figure}
\begin{center}
\epsfig{file=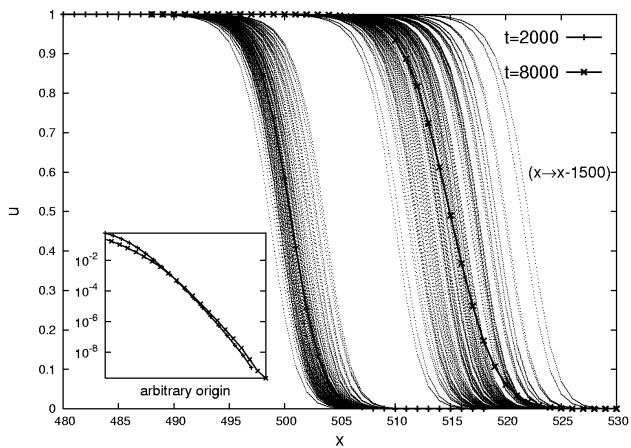,width=8.5cm}
\end{center}
\caption{\label{fig:disp}(From Ref.~\cite{EGBM2005}).
100 realizations of the evolution of a stochastic traveling wave
model, at two different times. 
The averages over the realizations (that corresponds to the physical
amplitude in QCD, see the dictionary~(\ref{dictionary})) 
are shown in full line.
{\it Inset:} 
the average $\langle u\rangle$ over the different realizations,
shifted in time and superimposed
in order to show the mismatch in slopes.
}
\end{figure}

The traveling wave 
that develops at large times
is characterized by its mean
shape and by the statistics of its
position $X_t$.
The latter may be defined, for example, as the value of $x$ for which $u$
has a definite value $u_0$, for instance $u_0=1/2$.
It is known 
since the seminal papers of Brunet and Derrida \cite{BD}
that for large times and large $N$, 
$X_t$ has the following mean and variance:
\begin{equation}
\begin{split}
&\langle X_t\rangle =\left(
\frac{\chi(\gamma_0)}{\gamma_0}
-\frac{\pi^2\gamma_0\chi^{\prime\prime}(\gamma_0)}
{2\ln^2 N}\right)t\\
&\langle X_t^2\rangle-\langle X_t\rangle^2 \sim \frac{t}{\ln^3 N}.
\end{split}
\label{stat}
\end{equation}
The average is taken over many realizations of the evolution~(\ref{evolgen}),
and $\chi(\gamma)$ is the characteristic function of the diffusion kernel
(i.e. the eigenvalue of $\chi(-\partial_x)$ 
that corresponds to the eigenfunction
$e^{-\gamma x}$).
$\gamma_0$ solves $\chi(\gamma_0)=\gamma_0\chi^\prime(\gamma_0)$. 
The shape of each realization
is, up to some noise
\begin{equation}
u(t,x)=e^{-\gamma_0(x-X_t)}.
\label{frontshape}
\end{equation}
A numerical evolution of a stochastic model of the F-KPP type
is displayed in Fig.~\ref{fig:disp}. One sees the universality in
shape of each individual realization as well as the dispersion in
front positions between different realizations, that indeed
grows like $\sqrt{t}$ (see Eq.~(\ref{stat})).
We will come back to the derivation of these results 
in Sec.~\ref{lecture2}.

%%%%%%

\subsection{Evolution of QCD amplitudes}

With the intuition gained by studying simple particle models,
we are now in position to argue that high energy scattering
naturally belongs to the universality class of the stochastic 
F-KPP equation.

\begin{figure*}
\begin{center}
\epsfig{file=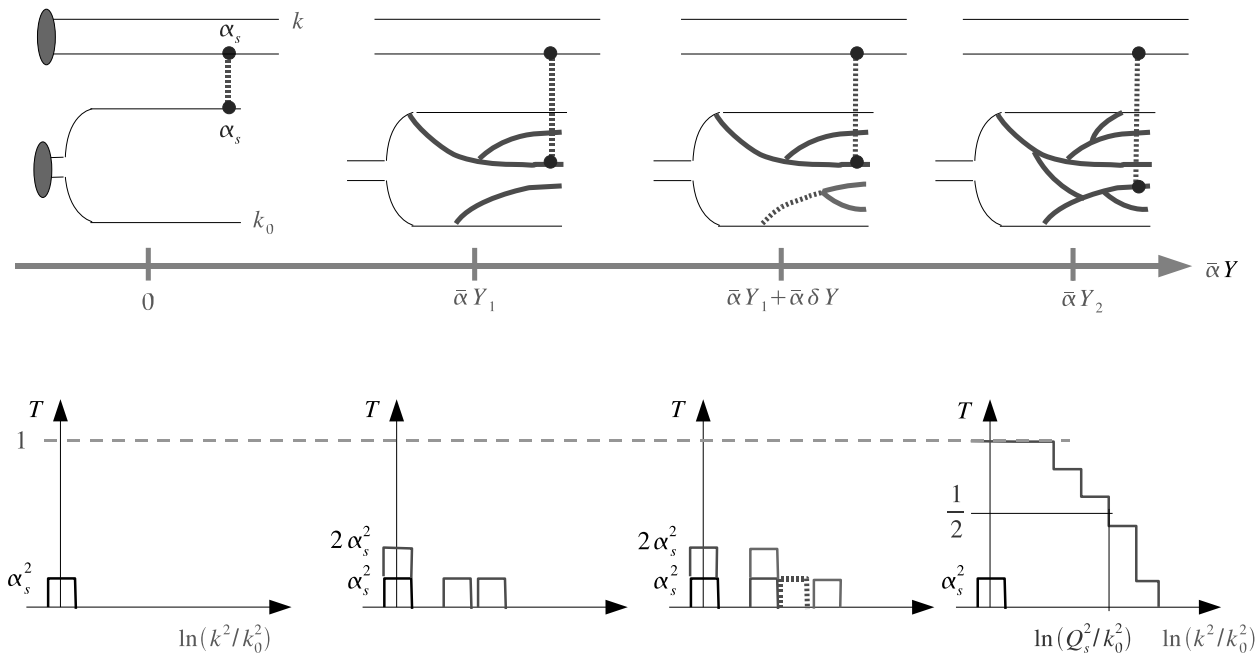,width=17cm}
\caption{\label{fig:QCD}One realization of the evolution 
of an onium-onium scattering amplitude.
At zero rapidity (leftmost plots), the amplitude consists in a bare gluon
exchange
between the probe (of size $1/k$; upper bubble) 
and the target (of size $1/k_0$; lower bubble):
$T=\alpha_s^2$ if the momentum scales of the two objects match, $T=0$ otherwise.
When rapidity is increased (going from the left to the right),
each gluon in the Fock state of the target may split in two gluons of
similar transverse momenta, and this process drives an exponential increase of
the amplitude for $k$ around $k_0$, up to a diffusion.
Eventually (rightmost plots), this growth has to slow down when 
$T\sim 1$ in order to comply with the 
unitarity constraints.
Beyond this point, nonlinear saturation effects enter and
a traveling wave forms in $T$. 
}
\end{center}
\end{figure*}

We consider the scattering of two hadrons made of 
quark-antiquark pairs
of respective sizes $1/k$ and $1/k_0$,
to simplify the discussion (see Fig.~\ref{fig:QCD}).
We assume that these hadrons are small enough (much smaller than
$1/\Lambda_\text{QCD}$) so that perturbation
theory applies.

If there is few energy 
available for the scattering
(the hadrons interact almost at rest), 
then typically they are in their valence configuration.
Their interaction consists in exchanging one gluon when they are of similar
size and at the same impact parameter: In this case, the forward
elastic amplitude reads $T\sim\alpha_s^2$. 
If the hadrons are of very different sizes or
if their impact parameters do not match, then they do not ``see'' 
each other and $T=0$.
Now boost one of the hadrons, called the target: This corresponds to increasing
the center-of-mass energy of the reaction. Due to the resulting
opening 
of phase space,
a typical configuration of the target is no longer the bare valence
quarks, but the latter together with a number $n(Y,k)$ of gluons of 
transverse momentum $k$, whose mean depend
on the rapidity. Correspondingly, the scattering amplitude off that particular
configuration of the target reads 
\begin{equation}
T(Y,k)\sim\alpha_s^2 n(Y,k),
\label{Tn}
\end{equation}
where $k$ is of the order of the inverse
size of the hadronic probe.
This relationship is not exact, but may be derived rigorously as a limit of
QCD scattering amplitudes.\footnote{%
Actually, it is useful to view the interaction
in the framework of the color dipole model \cite{M1994}, in
which gluons are assimilated to zero-size
quark-antiquark pairs whose components combine with
each other in the form of colorless dipoles, that evolve independently when
rapidity is increased. $n$ is in fact the number of such dipoles.}

In a given realization of the Fock space of the
target, the number of gluons typically doubles
when $\bar\alpha Y$ is increased by one unit
(each gluon has a probability to split in 2 gluons). 
The transverse momenta of the
new gluons are close to the ones of their parents, up to some diffusion.
This branching diffusion process is encoded in the BFKL evolution kernel
$\chi$ defined in the Introduction.
As we are following one given realization, 
that evolves randomly,
the equation contains a stochastic term of order $\sqrt{n}$, i.e. 
$\alpha_s\sqrt{T}$ from Eq.~(\ref{Tn}).
At this stage, we may write the evolution of the scattering amplitude off one
particular realization as
\begin{equation}
\partial_{\bar\alpha Y}T=\chi(-\partial_{\ln k^2})T+\alpha_s\sqrt{2T}\nu
\label{linearQCD}
\end{equation}
where $\nu$ is a noise of zero-mean whose variations are 
of order one when $\ln k^2$ and $\bar\alpha Y$ are varied by one unit.

Because $\tilde T$ is related to an interaction probability, 
$\tilde T\leq 1$ in appropriate
normalizations and so
the number of gluons of given momentum
(related to $T$) cannot grow exponentially forever. Hence a negative term
that becomes important when $T\sim 1$ has to enter Eq.~(\ref{linearQCD}).
One may write \cite{M2004,M2005,EGBM2005}
\begin{equation}
\partial_{\bar\alpha Y}T=\chi(-\partial_{\ln k^2})T-T^2+\alpha_s\sqrt{2T}\nu.
\label{QCD}
\end{equation}
The physical amplitude $A$ is obtained from $T$ by taking the average 
over all possible Fock realizations of the target, $A=\langle T\rangle$.
The mean-field approximation $\langle T^2\rangle=A^2$
of this equation precisely coincides with the BK equation~(\ref{BK}).

Hence we are led to an equation in the same universality class 
as Eq.~(\ref{evolgen}),
with the following dictionary (compare Eq.~(\ref{evolgen}) 
and Eq.~(\ref{QCD})):
\begin{equation}
\begin{split}
&u=T,\ 
t=\bar\alpha Y,\ x=\ln k^2,\\ 
&N=\frac{1}{\alpha_s^2},\ 
\chi(-\partial_x)=\text{BFKL kernel}.
\end{split}
\label{dictionary}
\end{equation}

Let us comment on the physical implications of having a 
stochastic equation
like~(\ref{QCD}) describing the dynamics.
The BK equation~(\ref{BK}) is known to admit traveling wave
solutions~\cite{MP2003}, whose position $X_t$ (which corresponds
in QCD
to the so-called saturation
scale $Q_s(Y)$ through $X_t=\ln Q_s^2(Y)$) 
is a deterministic function of the rapidity.
This property is related to a phenomenon observed in the
deep-inelastic scattering data and
called ``geometric scaling'' \cite{SGK2001}: 
The physical
amplitude $A(Y,k)$ only depends on the combined variable
$k^2/Q_s^2(Y)$, and not on $k$ and $Y$ separately.
By contrast, each realization of an evolution given by Eq.~(\ref{QCD}) is a
noisy traveling wave, which has a random position (i.e. saturation scale) 
whose statistics follow Eq.~(\ref{stat}) (use the
dictionary~(\ref{dictionary})).
The consequence is that $A=\langle T\rangle$ has a peculiar
scaling behavior, different from geometric scaling \cite{M2004,M2005,EGBM2005}:
\begin{equation}
A(Y,k)=A\left(\frac{\ln k^2
-\bar\alpha Y\left(\frac{\chi(\gamma_0)}{\gamma_0}
-\frac{\pi^2\gamma_0\chi^{\prime\prime}(\gamma_0)}{2\ln ^2(1/\alpha_s
  ^2)}
\right)
}{\sqrt{\bar\alpha Y/\ln^3 (1/\alpha_s^2)}}\right),
\label{solqcd}
\end{equation} 
which is easy to derive from Eqs.~(\ref{stat}),(\ref{frontshape}).

Mueller and Shoshi were the first authors who noticed
that geometric scaling was broken beyond the BK equation.
However, the statistical interpretation given here was crucial to
arrive at the correct scaling form~(\ref{solqcd}) 
(see Ref.~\cite{M2004}).

Let us summarize the assumptions that have lead to Eqs.~(\ref{QCD})
and~(\ref{solqcd}).
The linear terms therein represent the perturbative splitting of the
partons when rapidity increases, and may be obtained exactly
e.g. in the framework of the color dipole model \cite{M1994}.
We have not written down
explicitely the distribution of the 
noise $\nu$ (we just gave its typical
variations), but the solution~(\ref{solqcd})
is robust with respect to its detailed form.
The nonlinear term is supposed to encode parton saturation.
So far, we cannot provide an exact expression for it because a QCD
realization of saturation has not been derived.
However, the solution~(\ref{solqcd}) is also universal
with respect to the form of the
nonlinearities. The only really important condition
for universality arguments to apply is that
there should indeed be saturation, 
i.e. the gluon density should not
be able to grow forever. Arguments for this feature
were given a long time ago
\cite{GLRMQ}. So if the latter is true, because of the universality of
some properties of solutions to Eq.~(\ref{QCD}),
then we believe that Eq.~(\ref{solqcd}) 
represents the exact asymptotics of QCD.

Last, we would like to draw the attention of the reader to the fact
that we do not think that these results 
are very relevant phenomenologically yet, since
the asymptotics show up only for $\ln 1/\alpha_s^2\gg 1$ (which
means that $\alpha_s$ has to be unrealistically small). In our
opinion, one may conduct sound
phenomenological studies only once 
subleading effects have also been understood, beyond Eq.~(\ref{solqcd}).
However, numerical calculations 
may already provide some clues on the behavior of QCD amplitudes
for more realistic values of the parameters,
as explained in Ref.~\cite{M2006}.

%%%%%%%%%%%%%%%%%%%%%%%%%%%%%%%%%%%%%%%%%%%%%%%%%%%%%%%%%%%

\section{\label{lecture2}New general results on traveling wave equations 
and their applications to QCD}

In the previous section, we have argued that the evolution of 
QCD amplitudes at high energy 
is governed by a nonlinear stochastic evolution equation
in the universality class of the stochastic 
Fisher and Kolmogorov-Petrovsky-Piscounov 
(F-KPP) equation.
The latter literally reads
\begin{equation}
\partial_t u=\partial^2_x u+u-u^2+\sqrt{\frac{2}{N}u(1-u)}\nu,
\label{sFKPP}
\end{equation}
where $\nu$ is a normal Gaussian noise, uncorrelated both in $t$ and in $x$.
The deterministic part of this equation 
(obtained by setting $\nu=0$)
was first written in 1937
in the context of studies of the spread of genes (or diseases) in a
population \cite{FKPP1937}. 
Only in 1983 
some properties of its solutions
were rigorously derived \cite{P2004}.
In particular, it was understood that 
the large-time solutions were traveling waves.
The dominant effect of the noise term 
in Eq.~(\ref{sFKPP})
on the shape and velocity of the wave front
was worked out by Brunet and Derrida in 1997. 
Note that the scaling form of the dispersion in the front position
had only been measured numerically.
We have recently been able to
make progress on
the statistics of the position of the front \cite{BDMM2005}.

In practice, we will discuss the more general equation~(\ref{evolgen}),
of which Eq.~(\ref{sFKPP}) is just a particular case.

%%%%%%

\subsection{More on the propagation of noisy traveling waves}

\subsubsection{Solving the deterministic F-KPP equation}

First, we address the deterministic version of Eq.~(\ref{sFKPP}),
namely
\begin{equation}
\partial_t u=\chi(-\partial_x) u-u^2.
\label{detFKPP}
\end{equation}
The choice
$\chi(-\partial_x)=\partial_x^2+1$ 
would correspond to Eq.~(\ref{sFKPP}) exactly, but
we keep the branching diffusion terms in a more
general form in order to be able to easily transpose the results 
to other contexts, such as the QCD evolution equations.

In the case of the deterministic F-KPP
equation,
there are two fixed points, $u=0$ and $u=1$
which are respectively unstable and stable under small perturbations.
Starting from a localized initial condition,
the branching diffusion
encoded in $\chi$
leads to a local exponential growth of $u$ and a
spread in $x$. When $u$ becomes close to 1, then the nonlinear term
tames the growth so that $u$ does not get bigger than 1.
This mechanism leads to a wave front that invades larger values of $x$ when
time flows, being pulled by its tail.

One may compute the asymptotic velocity of the front in a very simple way.
Indeed, because the wave front is ``pulled'' by its low density tail, where
the nonlinearity is negligible,
it is enough to solve the linear part of Eq.~(\ref{detFKPP})
to understand the properties of the wave front.

Consider a front decaying like $u_\gamma=e^{-\gamma(x-v(\gamma) t)}$, with $\gamma$
a given wave number and $v(\gamma)$ its velocity. Then it is clear from
Eq.~(\ref{detFKPP}) that
\begin{equation}
v(\gamma)=\frac{\chi(\gamma)}{\gamma}.
\label{velocityg}
\end{equation}
The general solution is a linear superposition of $u_\gamma$,
\begin{equation}
u(t,x)=\int d\gamma f(\gamma)e^{-\gamma(x-v(\gamma) t)}
\label{super}
\end{equation}
where $f(\gamma)$ is a representation of the initial condition.

Let us concentrate on the large-time behavior of the solution.
We are following the wave around a fixed value of $u$ 
(since the nonlinearity
forces $u$ to range between 0 and 1), so the large-$t$ limit has to be 
performed
in the frame of the wave defined by the change of variable
$x_\text{WF}=x-V_\infty t$,
where $V_\infty$ is the asymptotic velocity of the wave. 
After replacement of $x$ in 
Eq.~(\ref{super}),
the saddle point condition reads $V_\infty=\chi^\prime(\gamma)$. 
Matching this
expression with Eq.~(\ref{velocityg}), we find that the wave
number $\gamma_0$ that dominates asymptotically solves 
$\chi(\gamma_0)/\gamma_0=\chi^\prime(\gamma_0)$, i.e. $\gamma_0$ 
minimizes $v(\gamma)$. 
Thus the asymptotic form of the wave front reads
\begin{equation}
\begin{split}
u(t\rightarrow\infty,x)
\sim e^{-\gamma_0(x-v(\gamma_0)t)}\ \ &\text{ for }\ \
x\gg v(\gamma_0)t,\\
&\text{ with }\
v^\prime(\gamma_0)=0.
\end{split}
\label{front}
\end{equation}
(The complete discussion may be found in Ref.~\cite{MP2003}).
This is the large time solution, but the front shape and
velocity are also known at subasymptotic times.
Let $X_t$ be the position of the front, defined e.g. by
$u(t,X_t)=u_0$, where $u_0$ is a constant between 0 and 1.
Then if the initial condition is localized enough, 
the wave front has
the shape \cite{M2004}
\begin{equation}
u(t,x)\sim(x-X_t)\exp\left(-\frac{(x-X_t)^2}
{2\chi^{\prime\prime}(\gamma_0) t}\right)
e^{-\gamma_0(x-X_t)}
\label{transients}
\end{equation}
and the front velocity reads
\begin{equation}
v_t=\frac{dX_t}{dt}=\frac{\chi(\gamma_0)}{\gamma_0}-\frac{3}{2\gamma_0 t}.
\label{vt}
\end{equation}
The asymptotic velocity $V_\infty$ (first term)
is reached quite slowly.
One more term is known in this large-$t$ expansion \cite{MP2004}, but we will not
discuss it here.
We may comment that 
the velocity~(\ref{vt})
corresponds to the velocity of a front that has reached its asymptotic 
shape~(\ref{front}) only in the range
\begin{equation}
(x-X_t)^2<2 \chi^{\prime\prime}(\gamma_0) t.
\label{range}
\end{equation}
This is indeed the range in which the Gaussian 
factor in Eq.~(\ref{transients}) is close to a constant.

%%%

\subsubsection{Hacking the deterministic equation to simulate discreteness}

Now we go back to the original stochastic equations~(\ref{evolgen}) 
or~(\ref{sFKPP}).
Generally speaking, what we missed when we neglected the noise
is mainly 
the discreteness of $u$. Think of a particle model on a lattice:
$u$ can take the values $1/N$, $2/N$, $3/N$\ldots but not fractions of
these numbers, a feature which of course was completely neglected 
in Eq.~(\ref{detFKPP}), as can be seen on the solution~(\ref{front}).
For $u\gg 1/N$, this might not be a problem, but for $u\sim 1/N$,
neglecting discreteness may prove very wrong, especially since the 
propagation of the wave front is very sensitive to its tail.

Brunet and Derrida came up with 
the idea that this discreteness could be incorporated
back into the deterministic equation~(\ref{detFKPP})  
by simply adding a cutoff\footnote{%
This cutoff coincides with the one argued by Mueller and Shoshi, but the 
interpretation as being a consequence of discreteness of the number of partons
had not been appreciated in Ref.~\cite{MS2004}. Historically,
we understood the relationship between QCD and statistical physics 
\cite{M2004}
by looking for an interpretation of the Mueller-Shoshi results,
in connection with the work of Brunet and Derrida \cite{BD}.
}
that forbids the values of $u$ between 0 and $1/N$ \cite{BD}.
They solved an equation of the type\footnote{
Strictly speaking, the $\Theta$-function in Eq.~(28) should not be 
applied to the diffusion term. In the case of the F-KPP equation, one would rather write
\begin{equation*}
\partial_t u=\partial_x^2 u + (u-u^2)\Theta(u-1/N),
\end{equation*}
although there would still be problems with the very mathematical 
definition of
such an equation. Eq. (28) would be correct literally if $t$ were 
discrete with steps $\Delta t$ of order 1.
}

\begin{equation}
\partial_t u=[\chi(-\partial_x) u-u^2]\Theta(u-1/N)
\label{cutoff}
\end{equation}
and claimed that the solution matched the first order of the full equation
in a small-noise (large-$N$) expansion.

We may estimate the large-$t$ velocity $V_\text{BD}$ 
of the traveling wave solution of Eq.~(\ref{cutoff}).
Starting from some initial condition, the front evolves toward
the asymptotic shape 
$e^{-\gamma_0(x-X_t)}$ and
its velocity increases according to Eq.~(\ref{vt}). However, as soon as the 
asymptotic front extends down to values of $u\sim 1/N$, that is for
$x\sim X_t+L$, where
\begin{equation}
L=\frac{\ln N}{\gamma_0},
\label{L}
\end{equation}
this shape cannot extend any further 
because of the cutoff in Eq.~(\ref{cutoff}). 
Then also the front velocity cannot grow any longer.
According to Eq.~(\ref{range}), this
happens at time $t\sim L^2/(2\chi^{\prime\prime}(\gamma_0))$, 
time at which the
velocity reads $V_\text{BD}=V_\infty-3c\chi^{\prime\prime}/(\gamma_0 L^2)$
according to Eq.~(\ref{vt}).
$c$ is a factor of order 1 whose determination needs a
more accurate 
calculation \cite{BD}. The complete result reads
\begin{equation}
V_\text{BD}=V_\infty-\frac{\pi^2 \chi^{\prime\prime}(\gamma_0)}{2\gamma_0 L^2}
=\frac{\chi(\gamma_0)}{\gamma_0}
-\frac{\pi^2\gamma_0\chi^{\prime\prime}(\gamma_0)}{2\ln^2 N}.
\label{VBD}
\end{equation}
The philosophy behind this procedure is quite transparent.
As soon as there are a few particles on a spatial site, 
parton number fluctuations are
negligible and the subsequent evolution in that bin is deterministic. 
(One can get convinced of this statement already by looking
at realizations of the evolution in the zero-dimensional model,
see Fig.~\ref{fig:zeroD}).
On the other hand, in the bins in which 
the deterministic evolution predicts a number of particles
less than 1 ($u<1/N$), one just sets $u$ to 0
in order to simulate discreteness. 
This is the effect of the cutoff.

\subsubsection{Incorporating stochastic effects}

The Brunet-Derrida cutoff procedure led to the following result: 
The front propagates at a velocity $V_\text{BD}$
lower than the velocity predicted by the mean-field equation~(\ref{detFKPP}), 
and its shape is the decreasing exponential (Eq.~(\ref{front})),
down to the position $x_\text{tip}(t)=X_t^\text{BD}
+\ln N/\gamma_0$ ($X_t^\text{BD}=V_\text{BD}t$), 
at which it
is sharply cut off.
\begin{figure}
\begin{center}
\epsfig{file=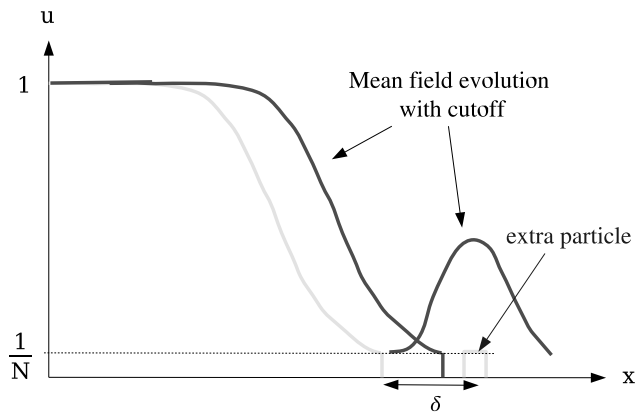,width=8.5cm}
\end{center}
\caption{\label{fluct}Evolution of the front 
with a forward fluctuation.
At time $t_0$, the primary front extends over a size 
$L$ (Eq.~(\ref{L}))
and is a solution of the Brunet-Derrida cutoff
equation~(\ref{cutoff}).
An extra particle has been stochastically generated at
a distance $\delta$ with respect to the tip of the
primary front.
At a later time, the latter grows deterministically into
a secondary front, that will add up to the primary one.
}
\end{figure}

But it may happen that a few extra 
particles are sent stochastically
ahead of the sharp tip of the front (See Fig.~\ref{fluct}).
Their evolution would pull the front forward.
To model this effect, we assume that the probability 
per unit time
that there be a particle
sent at a distance $\delta$ ahead of the tip simply 
continues the asymptotic
shape of the front, that is to say
\begin{equation}
p(\delta)=C_1 e^{-\gamma_0\delta},
\label{proba}
\end{equation}
where $C_1$ is a constant.
Heuristic arguments to support this assumption are presented
in Ref.~\cite{BDMM2005} (appendix~A).
Note that while 
the exponential shape is quite natural since it is the continuation
of the deterministic solution~(\ref{front}), 
the fact that $C_1$ need to be strictly
constant (and cannot be a slowly varying function of $\delta$)
is a priori more difficult to argue.

Now once 
a particle has been produced at position $x_\text{tip}+\delta$, 
say at time $t_0$,
it starts to multiply (see Fig.~\ref{fluct}) and
it eventually develops its own front (after a time of the order of $L^2$), 
that will add up to
the deterministic primary front made of
the evolution of the bulk of the particles.

Let us estimate the shift in the position of the front
induced by these extra forward particles.
Between the times $t_0$ and $t=t_0+L^2$,
the velocity of the secondary front is given by Eq.~(\ref{vt}).
Hence its position $X_t^{(2)}$,
after relaxation, will be given by
\begin{equation}
\begin{split}
X_t^{(2)}&=X_t^\text{BD}+\delta+\int_{t_0}^t dt^\prime\, v_{t^{\prime}-t_0}\\
&\sim X_t^\text{BD}+\delta-\frac{3}{2\gamma_0}\ln L^2 
\end{split}
\label{Xt2}
\end{equation}
where $X_t^\text{BD}=V_\text{BD}t$. Eq.~(\ref{Xt2}) holds
up to a constant. We have
used Eq.~(\ref{vt}) to express $v_{t^{\prime}-t_0}$.
The observed front will eventually result in the sum of the primary and
secondary fronts, after relaxation of the latter. 
Its position will be $X_t^\text{BD}$ supplemented by a shift
$R(\delta)$ that may be computed by writing the resulting front shape
as the sum of the primary and secondary fronts:
\begin{multline}
e^{-\gamma_0(x-X_t^\text{BD}-R(\delta))}=e^{-\gamma_0(x-X_t^\text{BD})}\\
+C_2 e^{-\gamma_0(x-X_t^\text{BD}-\delta+\frac{3}{2\gamma_0}\ln L^2)},
\label{sumfronts}
\end{multline}
where $C_2$ is an undetermined constant.
From Eq.~(\ref{sumfronts}) 
we can compute the shift 
due to the relaxation of a forward fluctuation $R(\delta)$:
\begin{equation}
R(\delta)=\frac{1}{\gamma_0}\ln\left(1+C_2\frac{e^{\gamma_0 \delta}}{L^3}\right).
\label{R}
\end{equation}
The probability distribution~(\ref{proba}) 
and the front shift~(\ref{R}) 
due to a forward fluctuation
define
an effective theory for the evolution of the position of the front $X_t$:
\begin{equation}
X_{t+dt}=\left\{
\begin{aligned}
&X_t+V_\text{BD} dt\ \ \text{proba}
\ \ 1-p(\delta)d\delta dt\\
&X_t+V_\text{BD} dt+R(\delta)\ \ \text{proba}
\ \ p(\delta)d\delta dt.
\end{aligned}
\right.
\end{equation} 
From these rules,
we may compute all cumulants of $X_t$:
\begin{equation}
\begin{split}
&V-V_\text{BD}=\int d\delta p(\delta)R(\delta)
=\frac{C_1C_2}{\gamma_0}\frac{3\ln L}{\gamma_0 L^3}\\
&\frac{[\text{$n$-th cumulant}]}{t}=\int d\delta p(\delta)[R(\delta)]^n\\
&\phantom{\frac{[\text{$n$-th cumulant}]}{t}} 
 =\frac{C_1C_2}{\gamma_0}\frac{n!\zeta(n)}{\gamma_0^n L^3}.
\end{split}
\label{cumulants}
\end{equation}

We see that the statistics of the position of the front still depend 
on the product $C_1 C_2$ of
the undetermined constants $C_1$ and $C_2$. We
need a further assumption to fix
its value.

We go back to the expression for the correction to
the mean-field front velocity, given in
Eq.~(\ref{cumulants}).
From the expressions of $R(\delta)$ (Eq.~(\ref{R})) and of
$p(\delta)$ (Eq.~(\ref{proba})), we see that
the integrand defining $V-V_\text{BD}$
is almost a constant function of $\delta$ for 
$\delta<\delta_0=3\ln L/\gamma_0$, and is decaying exponentially for
$\delta>\delta_0$. Furthermore, $R(\delta_0)$ is
of order 1, which means that when a fluctuation is emitted at
a distance $\delta\sim\delta_0$ ahead of the tip of the front,
it evolves into a front that matches in position the
deterministic primary front.
We also notice that when a fluctuation has $\delta<\delta_0$,
its evolution is completely linear until it is incorporated
to the primary front, whereas fluctuations with $\delta>\delta_0$
evolve nonlinearly but at the same time have a very 
suppressed probability.
We are thus led to the natural conjecture that the average 
front velocity is given by
Eq.~(\ref{VBD}), with the replacement 
$L\rightarrow \ln N/\gamma_0+\delta_0$.
The large-$N$ expansion of the new expression of the velocity
yields a correction  of the order of
$\ln\ln N/\ln^3 N$
to the Brunet-Derrida result,
more precisely
\begin{equation}
V=\frac{\chi(\gamma_0)}{\gamma_0}
-\frac{\pi^2\gamma_0\chi^{\prime\prime}(\gamma_0)}{2\ln^2 N}
+\pi^2\gamma_0^2\chi^{\prime\prime}(\gamma_0)
\frac{3\ln\ln N}{\gamma_0\ln^3 N}.
\label{vcorr}
\end{equation}
Eqs.~(\ref{cumulants}) and~(\ref{vcorr}) match
for the choice $C_1 C_2=\pi^2\chi^{\prime\prime}(\gamma_0)$. 
From this determination of $C_1 C_2$, 
we also get the full expression of the
cumulants:
\begin{equation}
\frac{[\text{$n$-th cumulant}]}{t}=
\pi^2\gamma_0^2\chi^{\prime\prime}(\gamma_0)
\frac{n!\zeta(n)}{\gamma_0^n\ln^3 N}.
\label{cumcorr}
\end{equation}
We note that all cumulants are of order 
unity for $t\sim \ln^3 N$, which 
is the sign that the distribution of the front position is far from a trivial 
Gaussian, which makes it particularly interesting.
On the other hand, they are proportional to $\kappa=t/\ln^3 N$, 
which is the sign that
the position of the front is the result of the sum of $\kappa$ 
independent random variables,
and as such, becomes Gaussian when $\kappa$ is very large.

These results rely on a number of 
conjectures that no-one has been able to prove
so far. However, we performed very precise numerical checks on specific models,
and we found a perfect matching (see Ref.~\cite{BDMM2005}). So we are reasonably
confident that our expressions are the correct ones. Providing a formal
proof would be an interesting challenge for a mathematician.

The analytical results presented here,
namely Eqs.~(\ref{vcorr}) and~(\ref{cumcorr}), may directly 
be applied to QCD
using the dictionary~(\ref{dictionary}). This would help 
to improve the analytic form of
the amplitude~(\ref{QCD}).

%%%%%%%%%%%%%%%%%%%%%

\subsection{Traveling waves in evolution models}

\begin{figure}
\begin{center}
\epsfig{file=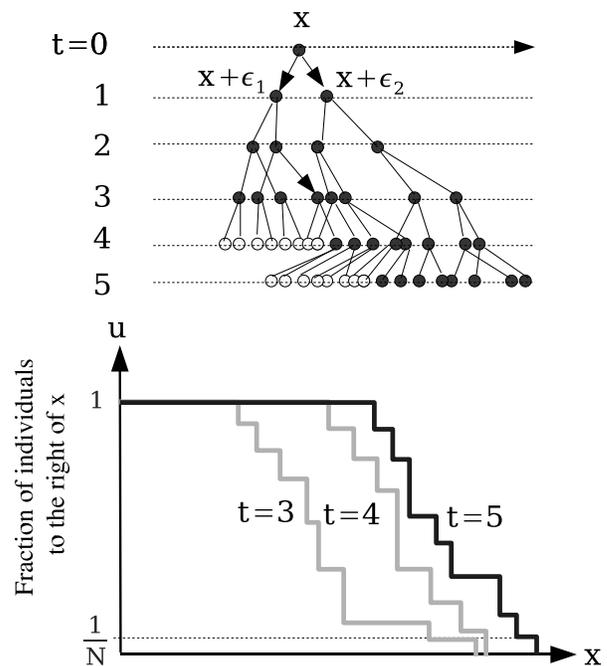,width=8cm}
\end{center}
\caption{\label{evol}Evolution of the population over a few generations.
In this model, each individual of a given generation $t$
has exactly two offspring, whose adequacy to environment differs from
his by the random variables $\varepsilon_1$ and $\varepsilon_2$.
Once the population size has reached $N$ ($N=8$ in this sketch), it
is kept constant by allowing only the ``best'' $N$ individuals to
have a descendance.
The fraction $u$ of individuals having their adequacy variable larger
than $x$ is plotted (bottom plot) for 3 different times. One sees that
$u$ exhibits the characteristics of
a traveling wave.
}
\end{figure}

Let us now consider a model of population evolution.
Each individual
is characterized by a single real number $x$ that measures its adequacy to
the environment.
At time $t$, there are $n(t,x)$ individuals in the population
with a given adequacy $x$.
To go from time $t$ to $t+1$, we give the following rule: Each individual
dies after having given birth to 2 offspring, that have respective adequacies
$x+\varepsilon_1$ and $x+\varepsilon_2$, where $\varepsilon_{1,2}$ are
random variables distributed according to a sufficiently local probability
distribution $\psi$.
If the total population exceeds a given integer $N$, then we get 
rid of the individuals
with the lowest values of $x$
in order to keep the population size constant and equal to $N$.
This model could represent for example the evolution of a population of bacterias
under asexual reproduction, in a medium where resources are limited, which enforces
a selection of the ``best'' individuals.
Let us define $u(t,x)$ to be the fraction of population that has 
its adequacy variable
larger than $x$. It is not difficult to realize that $u$ has the
shape of a wave front, 
connecting
a region where $u=1$ (for small values of $x$), 
and a region where $u=0$ (for larger values of $x$), that
moves toward positive $x$ when time elapses, see Fig.~\ref{evol}.

One may write a stochastic evolution equation for $u$:
%\begin{widetext}
\begin{multline}
u(t+1,x)=\min \left(1,2\int d\varepsilon\psi(\varepsilon)u(t,x-\varepsilon)\right)+\\
\sqrt{\frac{2}{N}\int d\varepsilon\psi(\varepsilon)u(t,x\!-\!\varepsilon)}
\nu(t+1,x)
\label{stochaevol}
\end{multline}
%\end{widetext}
where as usually, $\nu$ is a noise of zero mean and variance of order 1 
in the region where $u$ is small.
At first sight, it is far from obvious that this equation should 
be in the universality
class of the F-KPP equation. First, 
this is a finite difference equation in $t$ rather than
a differential equation. Second, the noise term is non-Gaussian
and furthermore, a closer look shows that $\nu$ is strongly correlated spatially.
The nonlinearity is also of completely different nature in this model.
However, the fact that a noisy front is formed for $u$
and especially the physical mechanism of evolution 
(branching diffusion+saturation)
points toward the F-KPP universality class.
If this is true, we can apply the results obtained in the previous
subsection to study how the mean adequacy $x$ grows with time 
(this would be the average position of the front).
It is easy to see that for this model, the characteristic function would be
\begin{equation}
\chi(\gamma)=\ln\left(2\int d\varepsilon\psi(\varepsilon)
e^{\gamma\varepsilon}\right).
\end{equation}
We checked that a numerical simulation of this model indeed matches
the analytical predictions, confirming that the model~(\ref{stochaevol}) is
indeed in the universality class defined by the F-KPP evolution.

For that kind of 
evolution models, 
one may ask a further question. At a given time $t$,
one may pick a given number of individuals, say 2 or 3, and count the number 
of generations
one has to go back in the past to find their most recent 
common ancestor. We denote these numbers
by $T_2$, $T_3$\ldots This question may be relevant to studies of the genetic
diversity of a population.

\begin{figure*}
\begin{center}
\epsfig{file=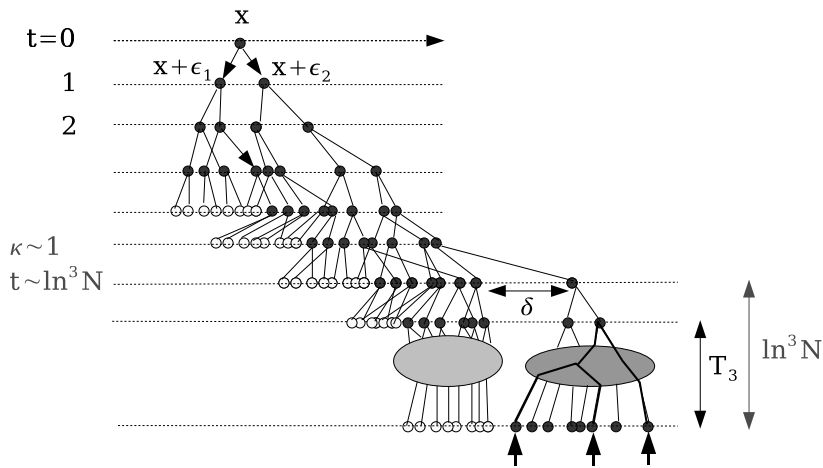,width=11cm}
\end{center}
\caption{\label{evol2}Evolution of the population over a few generations.
Each individual has two offspring in the next generation,
but only the $N$ ($=8$ in this illustration) with highest values of
$x$
are allowed to have a descendance. Sometimes, an individual
is so advanced 
(it has to be ahead by at least
$\delta=\delta_0=\frac{3}{\gamma_0}\ln\ln N$)
that his descendance will take over the 
whole population. This happens, 
on the average, once in $\ln^3 N$
generations. An estimator 
of this time scale may be obtained by picking 2, 3 or more
 individuals at random and looking for their common ancestor.
}
\end{figure*}

It is not difficult to guess the order of magnitude of $T_i$. 
We saw in the
previous subsection that $X_t$ looked like
the sum of $\kappa=t/\ln^3 N$ independent random variables.
One can understand this fact 
very precisely on this model. Indeed, from time to
time, the evolution generates an individual whose adequacy to
the environment is much larger than the one of any other individual. 
His offspring will partly inherit his adequacy, according to the stochastic
evolution rules and thus will still be in advance with respect to the
bulk of the population.
After some generations, only his descendants may survive, so that 
he will be the common ancestor of the whole population. 
For an individual to have a significant probability to have
his descendance replacing the whole population at some later time, 
the position of the secondary front developed
by this individual has to be larger than the position of the primary front,
a condition which reads $\delta>\ln L^3/\gamma_0$. 
According to Eq.~(\ref{proba}), this happens once
in $L^3=(\ln N/\gamma_0)^3$ generations. Hence
\begin{equation}
T_i\sim \ln^3 N.
\label{ctime}
\end{equation}
It turns out that through a slightly more elaborate analysis,
we may also fully compute the coefficients of the $T_i$, 
simply from the assumptions of the
previous subsection on the mechanism of the propagation of the front.
We found \cite{BDMM2006}
\begin{equation}
\frac{\langle T_3\rangle}{\langle T_2\rangle}=\frac{5}{4},\ \ \ 
\frac{\langle T_4\rangle}{\langle T_2\rangle}=\frac{25}{18},\ \cdots
\label{coalescence}
\end{equation}
These numbers characterize the statistics of the genealogical trees
that can be drawn in this kind of models.

It is not clear if results such as~(\ref{ctime}) and~(\ref{coalescence})
may be applied to particle physics observables.
The latter may be more relevant to evolutionary biology.

%%%%%%%%%%%%%%%%%%%%

\subsection{Spin glasses}

A spin glass \cite{sg} is a system of spins with random interactions.
The Hamiltonian is given by Eq.~(\ref{Hising}), but the
$J_{ij}$ are now random numbers 
endowed with a given probability distribution.
The simplest case is when $J_{ij}$ are binary variables $\pm 1$,
but the standard choice is a Gaussian distribution which leads
to a model named after Edwards and Anderson. The
Sherrington-Kirkpatrick model
is the infinite-range version of the latter: The same distribution
holds for any pair of spins and not only for nearest neighbors.

\begin{figure}
\begin{center}
\epsfig{file=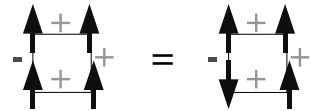,width=4cm}
\end{center}
\caption{\label{fig:spinglass}Two frustrated configurations of
4 spins, that have the same energy. 
The ``$+$'' and ``$-$'' on the lattice links
represent the values of the interaction$J_{ij}$.}
\end{figure}

This time, the minimum energy state 
at zero temperature is not unique, contrarily to the Ising case
for which the minimum energy configuration is reached when all spins
are aligned.
This is due to ``frustration'' (the fact that all interactions can
never
be satisfied simultaneously), and is illustrated in
Fig.~\ref{fig:spinglass}.

To classify the configurations,
one may define the overlap between two spin configurations 
$a$ and $b$ as a function
that reflects the fraction of spins that are common to $a$ and $b$. 
The usual definition is
$q_{ab}=\frac{1}{N}\sum_i S_i^a S_i^b$.
Speaking through the hat,
the configurations that minimize the energy
are organized ultrametrically with respect to a distance 
derived from the
overlap. This means that the configurations may be represented 
as the leaves of a (random)
tree. The statistics of this tree, first found by Parisi \cite{sg2}, 
turn out to be exactly the same as the ones
of our genealogic trees in the previous subsection that are
encoded in the ratios~(\ref{coalescence}).

It should be clear to the reader that there is no a priori reason 
for such a feature.
The method for finding the statistics in Parisi's theory relies on the replica method
(see Ref.~\cite{BDMM2006} for details),
and has nothing to do with the way we derived the result in our case.
Whether this link is deep or accidental is an outstanding open
question,
for which we have no clue.

%%%%%%%%%%%%%%%%%%%%%%%%%%%%%%%%%%%%%%%%%%%%%%%%%%%%%%%%%%%

\section{Conclusion}

When it was first suggested that high energy QCD has something to do
with reaction-diffusion processes and that this
analogy leads to new quantitative predictions \cite{M2004}, 
many experts were skeptical and reluctant
to accept this new point of view.
Getting these ideas through has involved some passionate debates, 
to say the least.
But at the time of the present lectures, that is two years
later, this proposal has undoubiously triggered a burst of activity, 
that has spread in
many different directions.

The main advantage of viewing high energy scattering in the peculiar 
way that we have explained here is that it
provides a simple picture 
endowed with a clear physical interpretation. This contrasts with the
high technicity of
evolution equations for QCD amplitudes that had been derived 
over the last 10 years
such as the Balitsky hierarchy~(\ref{B}).
Furthermore, this picture has a direct and intuitive connection 
with the basics of the parton model.
This approach has paved the way for new results in QCD,
obtained from tools known in statistical physics.

Obviously, there is still room for improvements of our understanding
of high energy scattering.
Admittedly, our proposal relies on a few conjectures that will have
to be proved. In our opinion, 
the most problematic one is the assumption that the number of gluons
or of color dipoles saturates at $1/\alpha_s^2$ 
per unit of transverse phase space.
Of course, this has been part of the ``folklore'' 
of high energy physics 
at least since 
the seminal work of Gribov, Levin, Ryskin in the 80's \cite{GLRMQ}. 
However,
no QCD realization of saturation has yet been found. 
A relevant question would be, 
for example, how parton recombination may be
implemented in the framework of the color dipole model. 
Some interesting ideas have recently been put forward 
(see e.g. Ref.~\cite{AGL}), 
but a rigorous proof in QCD still has to be provided.

In the second lecture, 
we have presented some recent advances in statistical physics
to which we have been able to contribute.
Here also, there is an intriguing correspondence between two seemingly
different physical problems: stochastic
fronts and the theory of spin glasses.
Whether this is something deep or simply 
accidental still remains to be understood.\\

\begin{acknowledgments}
This work was supported in part by the French-Polish research program
POLONIUM, contract 11562RG, 
and by the ECO-NET program, contract 12584QK.
\end{acknowledgments}

%%%%%%%%%%%%%%%%%%%%%%%%%%%%%%%%%%%%%%%%%%%%%%%%%%%%%%%%%%%


\begin{thebibliography}{00}

\bibitem{B}
I. Balitsky, Nucl. Phys.{\bf  B463} (1996) 99;
%%CITATION = HEP-PH 9509348;%%
  Phys. Rev. Lett. {\bf 81} (1998) 2024;
%%CITATION = HEP-PH 9807434;%%
  Phys. Lett. {\bf B518} (2001) 235.
%%CITATION = HEP-PH 0105334;%%
 
\bibitem{K}
Y.V. Kovchegov, Phys. Rev. {\bf D60} (1999) 034008;
%%CITATION = HEP-PH 9901281;%%
Phys. Rev. {\bf D61} (2000) 074018.
%%CITATION = HEP-PH 9905214;%%

\bibitem{LT}
  E.~Levin and K.~Tuchin,
  %``Solution to the evolution equation for high parton density QCD,''
  Nucl.\ Phys.\ B {\bf 573}, 833 (2000).
%  [arXiv:hep-ph/9908317].
  %%CITATION = HEP-PH 9908317;%%

\bibitem{MT2002}
  A.~H.~Mueller and D.~N.~Triantafyllopoulos,
  %``The energy dependence of the saturation momentum,''
  Nucl.\ Phys.\ B {\bf 640}, 331 (2002).
%  [arXiv:hep-ph/0205167].
  %%CITATION = HEP-PH 0205167;%%

\bibitem{MP2003}
S. Munier and R. Peschanski,
Phys. Rev. Lett. {\bf 91} (2003) 232001;
%%CITATION = HEP-PH 0309177;%%
Phys. Rev. {\bf D69} (2004) 034008.
%%CITATION = HEP-PH 0310357;%%

\bibitem{MP2004}
  S. Munier and R. Peschanski,
  Phys.\ Rev.\ D {\bf 70}, 077503 (2004).
  %%CITATION = HEP-PH 0401215;%%

\bibitem{GLRMQ}   L.V. Gribov, E.M. Levin and M. G. Ryskin, Phys. Rep.
{\bf 100} (1983) 1;
%%CITATION = PRPLC,100,1;%%
  A.~H.~Mueller and J.~w.~Qiu,
  %``Gluon Recombination And Shadowing At Small Values Of X,''
  Nucl.\ Phys.\ B {\bf 268}, 427 (1986).
  %%CITATION = NUPHA,B268,427;%%

 \bibitem{BFKL}
L.~N. Lipatov,
Sov. J. Nucl. Phys. {\bf 23}, 338 (1976);
%%CITATION = SJNCA,23,338;%%
E.~A. Kuraev, L.~N. Lipatov, and V.~S. Fadin,
Sov. Phys. JETP {\bf 45}, 199 (1977);
%%CITATION = SPHJA,45,199;%%
I.~I. Balitsky and L.~N. Lipatov,
Sov. J. Nucl. Phys. {\bf 28}, 822 (1978).
%%CITATION = SJNCA,28,822;%%

\bibitem{GBW}
  K.~Golec-Biernat and M.~W\"usthoff,
  %``Saturation effects in deep inelastic scattering at low Q**2 and its
  %implications on diffraction,''
  Phys.\ Rev.\ D {\bf 59}, 014017 (1999);
%  [arXiv:hep-ph/9807513].
  %%CITATION = HEP-PH 9807513;%%
  Phys.\ Rev.\ D {\bf 60}, 114023 (1999).
%  [arXiv:hep-ph/9903358].
  %%CITATION = HEP-PH 9903358;%%

\bibitem{MSM}
  S.~Munier, A.~M.~Sta\'sto and A.~H.~Mueller,
%   ``Impact parameter dependent S-matrix for dipole proton scattering from
  %diffractive meson electroproduction,''
  Nucl.\ Phys.\ B {\bf 603}, 427 (2001).
%  [arXiv:hep-ph/0102291].
  %%CITATION = HEP-PH 0102291;%%

\bibitem{MS2004}   A. H. Mueller and A. I. Shoshi, 
Nucl. Phys. B {\bf 692} (2004) 175.
  %%CITATION = HEP-PH 0402193;%%
 
\bibitem{M2004}
  S.~Munier,
  Proceedings of the Theory Summer Program on RHIC physics, BNL-73263-2004,
  Brookhaven, July 2004.

\bibitem{M2005}
  E.~Iancu, A.~H.~Mueller and S.~Munier,
  %``Universal behavior of QCD amplitudes at high energy from general tools  of
  %statistical physics,''
  Phys.\ Lett.\ B {\bf 606}, 342 (2005);
  %%CITATION = HEP-PH 0410018;%%
   S.~Munier,
  %``High energy scattering in QCD as a statistical process,''
  Nucl.\ Phys.\ A {\bf 755}, 622 (2005).
  %%CITATION = HEP-PH 0501149;%%


\bibitem{EGBM2005}
  R.~Enberg, K.~Golec-Biernat and S.~Munier,
  %``The high energy asymptotics of scattering processes in QCD,''
  Phys.\ Rev.\ D {\bf 72}, 074021 (2005).
   %%CITATION = HEP-PH 0505101;%%

\bibitem{M2006}
   S.~Munier, Phys.\ Rev.\ D {\bf 75} 034009 (2007).
  %%CITATION = HEP-PH 0608036;%%

\bibitem{SX2005}
  A.~I.~Shoshi and B.~W.~Xiao,
  %``Pomeron loops in zero transverse dimensions,''
  Phys.\ Rev.\ D {\bf 73}, 094014 (2006).
  %%CITATION = HEP-PH 0512206;%%

\bibitem{FKPP1937}
R.~A. Fisher,
Ann. Eugenics {\bf 7}, 355 (1937);
%%CITATION = ANEUA,7,355;%%
A.~Kolmogorov, I.~Petrovsky, and N.~Piscounov,
Moscou Univ. Bull. Math. {\bf A1}, 1 (1937).
%%CITATION = NONE;%%

\bibitem{P2004} For a recent review on stochastic fronts, 
see D. Panja, Phys. Rept. {\bf 393}
(2004) 87.
%%CITATION = PRPLC,393,87;%%

\bibitem{BD}
E. Brunet and B. Derrida, Phys. Rev. {\bf  E56} (1997) 2597;
%%CITATION = COND-MAT 0005362;%%
Comp. Phys. Comm. {\bf 121-122} (1999) 376;
%%CITATION = CPHCB,121,376;%%
J. Stat. Phys. {\bf 103} (2001) 269.
%%CITATION = JSTPB,103,269;%%
 
\bibitem{M1994} A.~H. Mueller, Nucl. Phys. B {\bf 415} (1994) 373.
%%CITATION = NUPHA,B415,373;%%

\bibitem{SGK2001}   A. M. Sta\'sto, K. Golec-Biernat and J. Kwieci\'nski,
Phys. Rev. Lett. {\bf 86} (2001) 596.
%%CITATION = HEP-PH 0007192;%%

\bibitem{BDMM2005}
  E.~Brunet, B.~Derrida, A.~H.~Mueller and S.~Munier,
%   ``A phenomenological theory giving the full statistics of the position of
  %fluctuating pulled fronts,''
  Phys.\ Rev.\ E {\bf 73}, 056126 (2006).
  %%CITATION = COND-MAT 0512021;%%

\bibitem{BDMM2006}
  E.~Brunet, B.~Derrida, A.~H.~Mueller and S.~Munier,
  Europhys.\ Lett.\  {\bf 76}, 1 (2006);
%  ``Noisy traveling waves: effect of selection on genealogies,''
%  arXiv:cond-mat/0603160,
%to appear in Europhys. Lett.;
  %%CITATION = COND-MAT 0603160;%%
extended version in preparation.

\bibitem{sg}
For a review on experimental and theoretical aspects
of spin glasses, see e.g. K.~Binder, A.~P.~Young,
Rev. Mod. Phys. {\bf 58}, 801 (1986). 

\bibitem{sg2}
G. Parisi,
%The  order parameter for spin-glasses - function on the interval 0-1,
 J. Phys. A {\bf 13},  1101 (1980);
M.~M\'ezard , G.~Parisi, N.~Sourlas, G.~Toulouse, M.~A. Virasoro,
% Replica symmetry-breaking and the nature of the spin-glass phase,
 Journal de Physique {\bf 45}, 843-854 (1984).

\bibitem{AGL}
  E.~Avsar, G.~Gustafson and L.~Lonnblad,
  %``Energy conservation and saturation in small-x evolution,''
  JHEP {\bf 0507}, 062 (2005).
  %%CITATION = HEP-PH 0503181;%%


\end{thebibliography}
\end{document}